\documentclass[a4paper]{article}
\usepackage[papersize={210mm,297mm},total={145mm,228mm},centering,includeheadfoot]{geometry}
\usepackage{xcolor}
\definecolor{linkcolor}{RGB}{170,0,0} 
\definecolor{citecolor}{RGB}{44,160,46} 
\definecolor{urlcolor}{RGB}{0,51,153} 
\RequirePackage[linktocpage,pdfcreator=easychair.cls-3.4,colorlinks=True,urlcolor=urlcolor,citecolor=citecolor,linkcolor=linkcolor]{hyperref}
\usepackage{graphicx}
\graphicspath{{./}{./figs/}}
\usepackage{doc}
\usepackage[utf8]{inputenc}
\usepackage{listings}
\newcommand{\inFGScale}{0.43}
\newcommand{\inBpScale}{0.55}
\newcommand{\inst}[1]{{$^#1$}}
\newcommand{\email}[1]{\texttt{#1}}

\title{\vspace{-22mm}CryptoMiniSat Switches-Optimization for Solving Cryptographic Instances}
\author{A.-M. Leventi-Peetz\inst{1} \and O. Zendel\inst{1} \and W. Lennartz\inst{2} \and K. Weber\inst{2}}
\date{\small%
  $^1$Federal Office for Information Security,\\
  Godesberger Allee 185--189, DE-53175 Bonn, Germany\\
  \email{leventi-peetz@bsi.bund.de}
\\
  $^2$inducto GmbH,\\
  Karl-Wastl-Straße 13, DE-84405 Dorfen, Germany
}
\begin{document}
\maketitle
\begin{abstract}
Performing hundreds of test runs and a source-code analysis, we empirically identified improved parameter configurations for the CryptoMiniSat (CMS) 5 for solving cryptographic CNF instances originating from algebraic known-plaintext attacks on 3 rounds encryption of the Small AES-64 model cipher SR$(3, 4, 4, 4)$. We finally became able to reconstruct 64-bit long keys in under an hour real time which, to our knowledge, has never been achieved so far. Especially, not without any assumptions or previous knowledge of key-bits (for instance in the form of side-channels, as in~\cite{Mohamed2012algebraicSCA}). A statistical analysis of the non-deterministic solver runtimes was carried out and command line parameter combinations were defined to yield best runtimes which ranged from under an hour to a few hours in median at the beginning. We proceeded using an Automatic Algorithm Configuration (AAC) tool to systematically extend the search for even better solver configurations with success to deliver even shorter solving times. In this work we elaborate on the systematics we followed to reach our results in a traceable and reproducible way. The ultimate focus of our investigations is to find out if CMS, when appropriately tuned, is indeed capable to attack even bigger and harder problems than the here solved ones. For the domain of cryptographic research, the duration of the solving time plays an inferior role as compared to the practical feasibility of finding a solution to the problem. The perspective scalability of the here presented results is the object of further investigations.
\end{abstract}

\section{Introduction}
CryptoMiniSat\footnote{Developed by Mate Soos as an open source community project~\cite{CMS5-2017,Soos2009CMS}.}
offers a wide range of parameter settings to choose when
calling the solver and these parameters seem to sensitively influence
the search for a solution in case of cryptographic CNF (\emph{conjunctive normal form})
instances. We selected parameter combinations that especially affect the solver
runtime-behavior in the case of CNF instances, generated from algebraic
equations systems that represent \emph{known-plaintext attacks} (KPA) on
the Small AES-64 model cipher.
The KPA is an attack model of cryptanalysis
where the attacker has access to both the plaintext and its encrypted
version (ciphertext). Knowing the cryptographic algorithm, the goal of
the attack is to reconstruct the secret key from the text information.
AES is an iterated block cipher which, as the encryption mechanism,
repeats a set of state transformations for a number of rounds.
The input (plaintext) and the output of each and every transformation
is called \emph{state} and consists of a fixed number of bits.
AES makes use of both non-linear and linear
transformations, one being the round-key addition. The round keys are
generated from the secret key.
To perform the state transformations, the state or \emph{block} is divided
into \emph{words} which are arranged in a rectangular array, structured by
rows and columns. In AES words are eight bits long.
The nonlinear transformation operates on each word independently.
For the key recovery one first describes the encryption algorithm in the form of a Boolean MQ (multi quadratic) polynomial equation system of bit variables, as introduced by Courtois and Pieprzyk~\cite{courtois2002:algres}.
Inserting the known bits of the plaintext and cyphertext pairs one gets a non-linear system of equations, the solution of which delivers the secret key.
Each additional text pair adds another set of equations to the system but with the same key bit variables. The equations of the round key expansion enter only once.
This system can be tackled with algorithms based on Gröbner bases and SAT solvers or alternatively only with a SAT solver, which is the approach followed here.
Routinely, so named \emph{small scale variants} of the AES polynomial
system~\cite{cid2005:smallAES} are employed for tests in the cryptographic
community. Relevant to the usual variants are the following numbers:
%
\begin{itemize}\setlength{\itemsep}{-0.3ex}\renewcommand{\labelitemi}{--}
\item $n$ is the number of (encryption) rounds,
\item $r$ is the number of  \emph{rows} in the rectangular arrangement of the state,
\item $c$ is the number of \emph{columns} in the rectangular arrangement of the state,
\item $e$ is the size (in bits) of a word, normally 4 or 8.
\end{itemize}

AES would be considered as broken when the model for $(n, r, c, e) = (10, 4, 4, 8)$ has been solved and the corresponding 128-bit long key has been recovered.
However, already successes to recover 8-bit and 16-bit
long keys for very small AES model ciphers are reported in the literature, mainly 
in association with benchmarking of SAT solvers in comparison to one
another~\cite{charfi2014bsthesis,indroy2018mthesis}. Here the solution of the model system SR$(3, 4, 4, 4)$ is discussed using CMS.

The computations have been performed on both four socket AMD Opteron
6378 and two socket EPYC 7551 systems with 256 GB RAM using 31 threads per job
(64 respectively 128 would have been possible) and up to four jobs parallelly.
Details concerning the derivation of the MQ algebraic equations system of
the attack and its transformation to CNF will
not be further discussed here because the focus of this work
lies on the configuration of the solver. The interested reader can
find some of these information
in~\cite{leventipeetz2017Sbox}. CryptoMiniSat solves all cases of 2
rounds encryption, that is the SR$(2, 4, 4, 4)$ case, for the Small AES-64 model within seconds.
Key recovery from 3 rounds encryption can get successfully
accomplished with the solver running in default parameter
setting, however mostly within hours. The solution finding is subjected to distinct statistical
variations, due to the indeterministic behavior of the solver in
multi-thread operation mode. When an upper runtime-limit has
been set, it is a matter of chance whether the solver will find the
solution or not. Running the solver determininistically in single-thread
modus is out of the question because it takes infeasible long.
The parallely and asynchronously running solver threads complement each other by exchanging information and are indispensable for finding the solution in tolerable time. 
A systematical statistical investigation of the solver's
behavior for a multitude of cases helped us find solver
parameter combinations which enable key recovery for 3 rounds
AES-64 encryption, or solving the SR$(3, 4, 4, 4)$ case, within predictable time-intervals. 
We extended our efforts beyond the empirical parameter optimization by
employing an automatic algorithm configuration tool which we adapted
for the problem and applied it to find even better parameter settings.

This paper is organized in 5 parts as follows: In the first part we
give an overview of the size, format and the density of the CNF
instances which we have worked with. In the second part we present and
discuss runtime statistics of the solver in its default parameter
settings. In the third part the empirical parameter optimization
investigations and their results are presented and discussed. Due to
the indeterministic behavior of the solver in multi-thread operation
mode, certain changes in the source code have been undertaken, which
were seen as necessary in order to increase the significance of the
influence of changes in parameter settings to the benefit of the
generation of distinct results. These source-code changes will be
substantiated and the out of them resultant improvements of the solver
runtimes will be graphically demonstrated. In the fourth part of this
paper, the innovative results of an Automatic Algorithm Configuration for the
parameters of CryptoMiniSat which produced even better parameter
configurations
will be presented and discussed.
We conclude with a summary and description of further planned
investigations to optimize CMS.

\section{Classification of CNF-Instances}
The plaintexts were made of english words and spaces.
We have varied the number of the KPA text pairs, used for the instances
generation, in order to investigate also the influence of this number on
the solution runtime for instances otherwise created with the same
key. The number of text pairs measures the redundancy of information given
to the solver, as all texts are encoded with the same algorithm (the
same logic) and
the same key. The bigger the number of text pairs, the
bigger the number of variables and constraints in the resulting
instance to solve, so one shouldn't expect to be able to always get an advantage
by arbitrarily increasing the number of text pairs. However, for each key case there seems to exists an \emph{optimal} number of text pairs minimizing the solution time and that has to be discovered.
The number of text pairs varied here between 16 and 32 pairs.
It should be noted that these cryptographic CNF instances only possess one truth assignment by construction.
An overview of the parameters of the tested CNF instances is listed in
table~\ref{tab:nums}.
%
%
\begin{table}[htbp]
  \caption{\label{tab:nums}Some characteristic numbers of the utilized CNF instances. Instance tokens comprise \emph{$\langle$no.\ of rounds$\rangle$-$\langle$key token$\rangle$-$\langle$no.\ of text pairs$\rangle$}.}
  \vspace{1ex}
  \begin{minipage}{\textwidth}
  \centering
  \begin{tabular}{c|r|r|r|r|r}
    \hline
    Instance\footnote{\textbf{k3}: 0123456789abcdef; \textbf{k4}: 0101010101010101; \textbf{k6}: b25286f7d3e7b3e1} & Rounds & Text Pairs & Variables $L$ & Clauses $N$ & Density $N/L$ \\
    \hline
    3-k4-12  & 3 & 12 &  4096 & 1228120 & 299.8 \\
    3-k4-16  & 3 & 16 &  5376 & 1626619 & 302.6 \\
    3-k4-30  & 3 & 30 &  9856 & 3021481 & 306.6 \\
    3-k3-12  & 3 & 12 &  4096 & 1227940 & 299.8 \\
    3-k3-14  & 3 & 14 &  4736 & 1427202 & 301.4 \\
    3-k3-18  & 3 & 18 &  6016 & 1825395 & 303.4 \\
    3-k3-20  & 3 & 20 &  6656 & 2024600 & 304.2 \\
    3-k3-22  & 3 & 22 &  7296 & 2224391 & 304.9 \\
    3-k3-30  & 3 & 30 &  9856 & 3021481 & 306.6 \\
    3-k6s-20 & 3 & 20 &  6656 & 2025228 & 304.3 \\
    3-k6s-22 & 3 & 22 &  7296 & 2224391 & 304.9 \\
    3-k6s-24 & 3 & 24 &  7936 & 2424008 & 305.4 \\
    3-k6s-30 & 3 & 30 &  9856 & 3021481 & 306.6 \\
    4-k6s-30 & 4 & 30 & 13760 & 4447760 & 323.2 \\
    \hline
  \end{tabular}
  \end{minipage}
\end{table}
%

%
We varied also the \emph{quality}
of the encoding key. Experiments were performed with several different
keys, belonging to three different classes. As representative of each class,
three keys are listed below, with which the results presented here have
been produced. 
\begin{description}\setlength{\itemsep}{-0.3ex}
  \item[k4] '0101010101010101'; \emph{pathologic} or insecure
  \item[k3] '0123456789abcdef'; \emph{structured}
  \item[k6] 'b25286f7d3e7b3e1'; secure, random
\end{description}
All instances contain clauses of varying length and all
instances are of the type \emph{sparse} and without inclusion of
explicit XOR-Clauses.

\section{Runtime Statistics for CMS in Default-Setting}
%

The CMS threads work asynchronously and the order in which they exchange
information is unpredictable depending on external influences like the
operating system and administrative tasks running on the
computer. Similarly indeterministic and irreproducible is each and
every solver run and solution process.  This circumstance leads to the
result that repeated solver attempts to solve one and the same
instance under identical parameter configuration can deliver very
different runtimes, which renders the nature of statements about
average runtime measurements to a statistical one.
Boxplots\footnote{See for example \textsc{Wikipedia} Box plot
  \url{https://en.wikipedia.org/wiki/Box_plot}.} have been chosen as
proper statistical analysis method for the runtime measurements.

\begin{figure}[htbp]
  \centering
  \includegraphics[scale = \inBpScale]{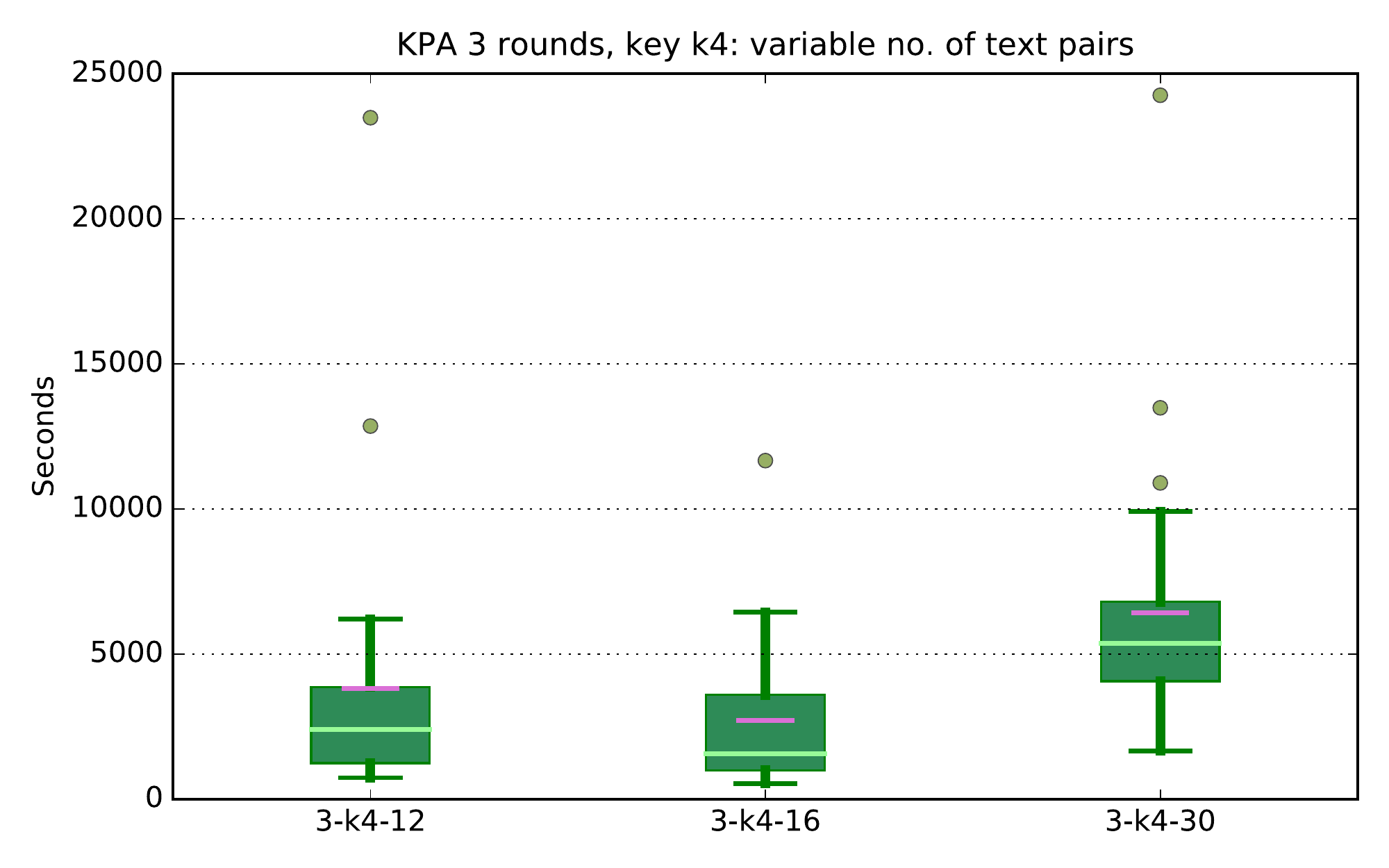}
  \caption[Boxplot 01]
  {\label{f:Boxplot01}3 rounds, key k4: varying number of text pairs.}
\end{figure}

\begin{figure}[htbp]
  \centering
  \includegraphics[scale = \inBpScale]{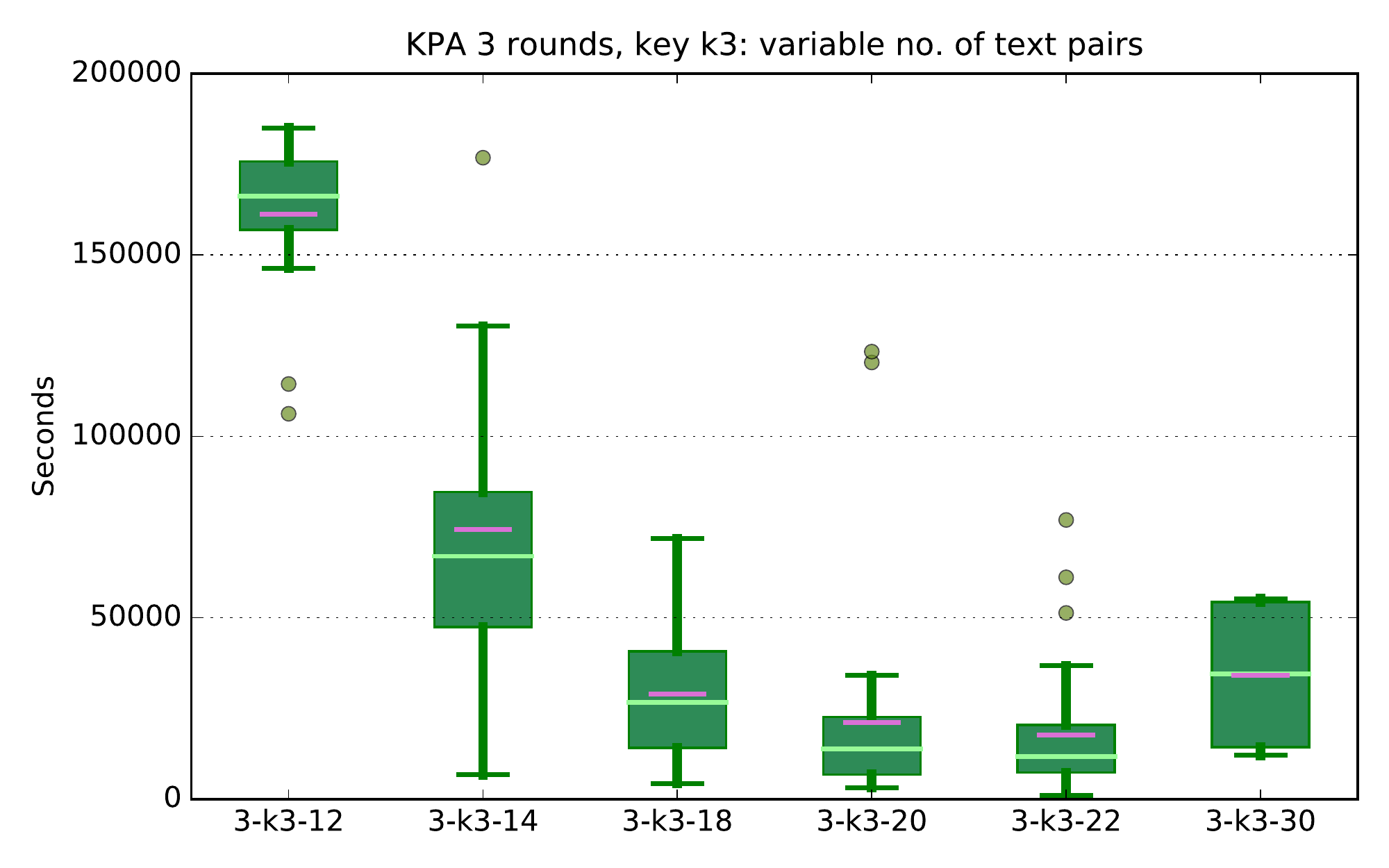}
  \caption[Boxplot 02a]
  {\label{f:Boxplot02a}3 rounds, key k3: varying number of text pairs.}
\end{figure}

\begin{figure}[htbp]
  \centering
  \includegraphics[scale = \inBpScale]{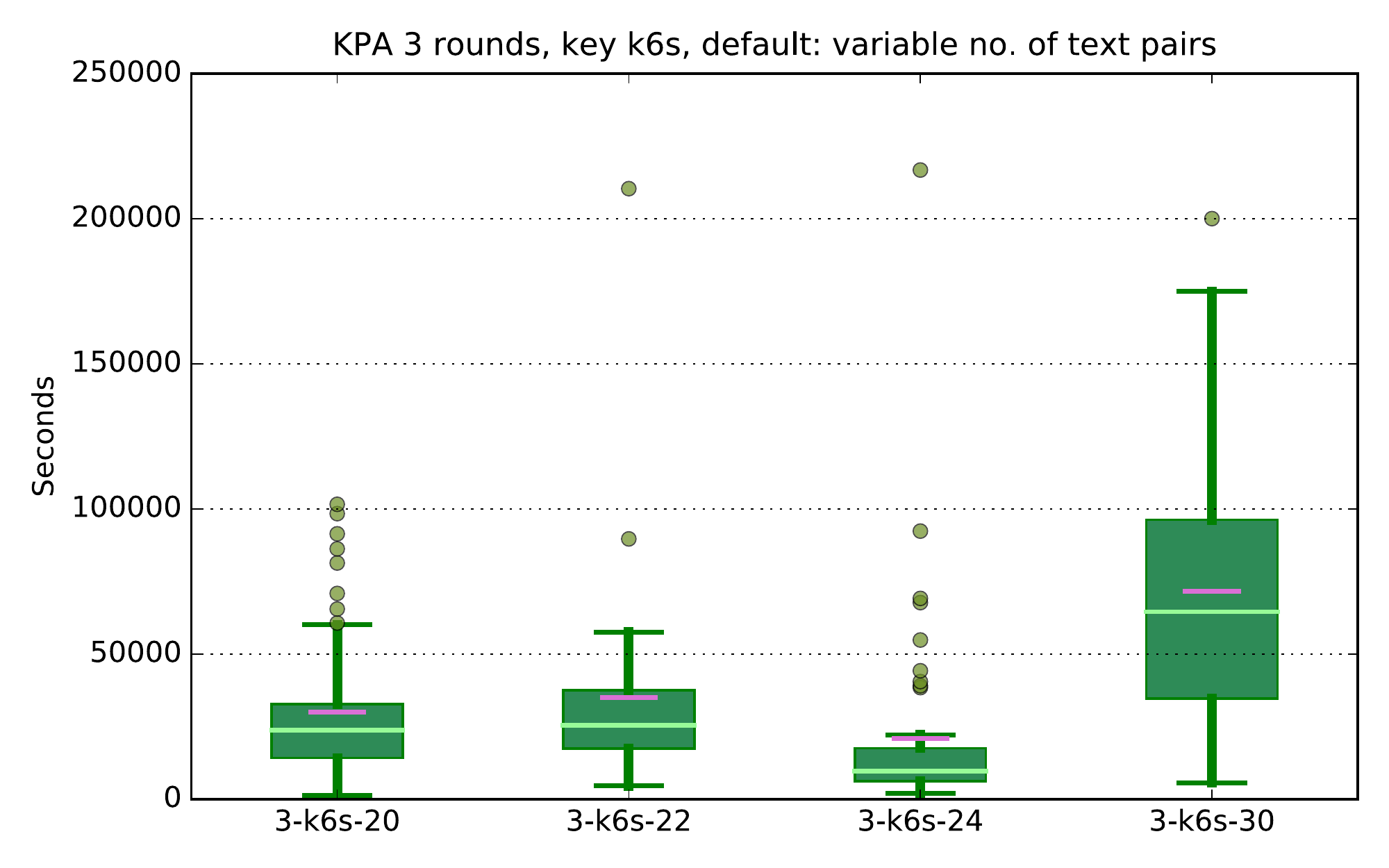}
  \caption[Boxplot 08a]
  {\label{f:Boxplot08a}3 rounds, key k6s: varying number of text pairs.}
\end{figure}

In Figure~\ref{f:Boxplot01} the runtime analysis of the solver for the
as \emph{insecure} classified key for 12, 16, and 30 text pairs
respectively is depicted. The median of the runtime varies with the
number of text pairs and the faster key recovery is achieved with the
instance created out of 16 text pairs.  Also the mean values of the
measured data reflect the same result though each assuming a higher
value than its respective median.
In Figure~\ref{f:Boxplot02a} there is depicted the runtime analysis of
the solver for the \emph{structured} key case and for 12, 14, 18, 20,
22, and 30 text pairs, respectively. Also in this case does the median
of the solution time distinctly vary dependent on the number of
text pairs, the optimal number appearing to be this time in the case
of 22 pairs. Again mean values and medians stay consistent to this
result with the mean values climbing a bit higher than the
medians. Obviously the use of the structured key makes the solution of
the problem considerably more expensive shifting the solution time one
order of magnitude towards higher values.
In plot~\ref{f:Boxplot08a} the solver runtime analysis regarding the
solution of instances created with the random or \emph{secure} key is
demonstrated.
The solution runtime is of the same order of magnitude as in the case
of the \emph{structured}, or simple key, the optimal number of text
pairs appearing to be 24 this time.  The CMS runtime measurements
produce a spectrum of random data representing solver
runtimes containing some few extremely long runtimes. However,
in each case the majority of the resulting values lie within a well
defined limited region.

\begin{figure}[htbp]
  \centering
  \includegraphics[scale = \inBpScale]{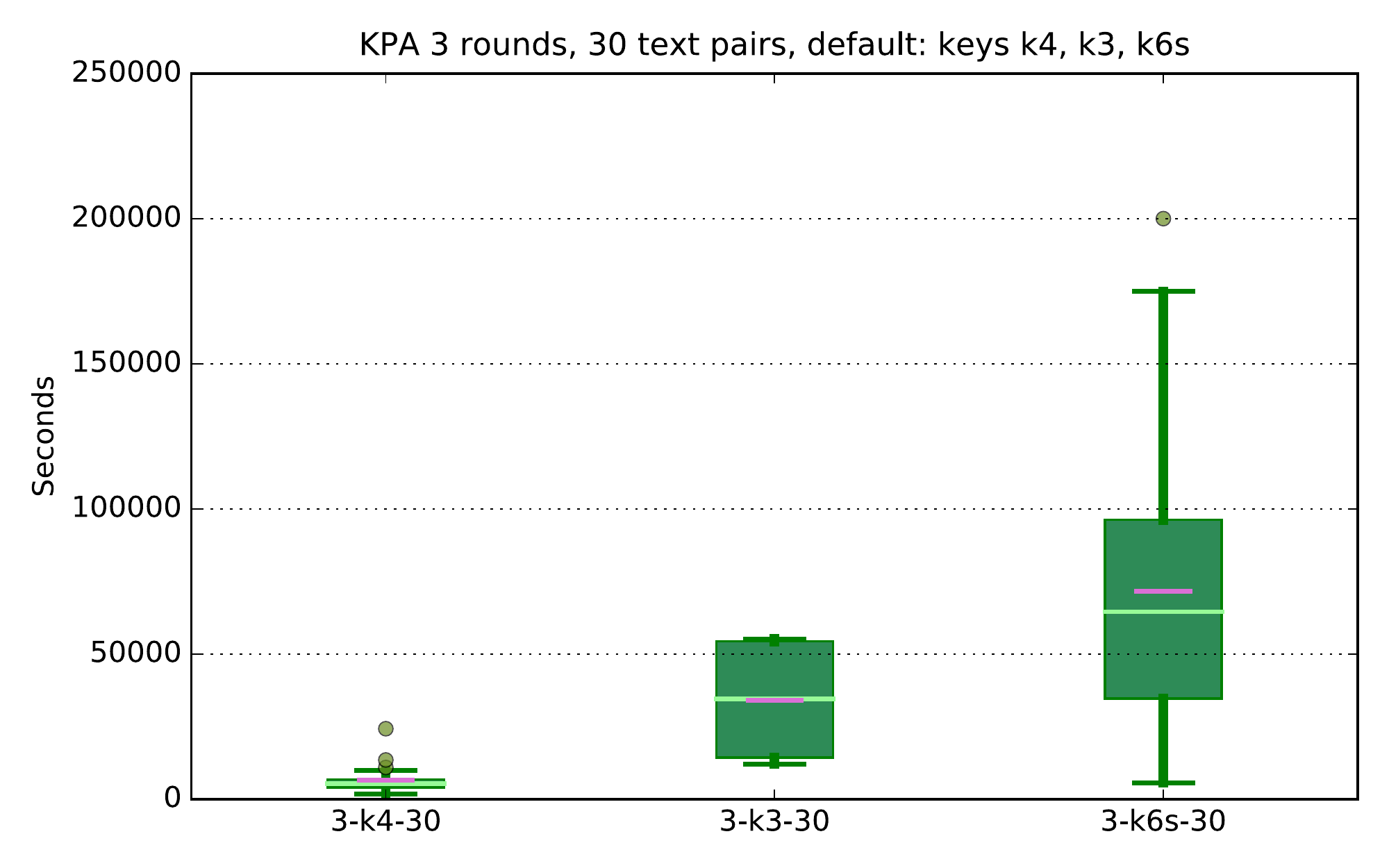}
  \caption[Boxplot 12a]
  {\label{f:Boxplot12a}3 rounds, constant number of text pairs,
    varying keys.}
\end{figure}

Comparing the solution times in the plots, one observes
that the number of text pairs is important, because
a convenient choice of this number can occasionally strongly diminish
the solver runtime for instances otherwise created with the same key.
Comparing solution runtimes for three different keys using
instances created with the same number of text pairs, one
can attest that a simple key costs a shorter solution runtime as
compared to the runtime needed to solve the instance
generated with a secure key, see Figure~\ref{f:Boxplot12a}.
In table~\ref{tab:runstats} the runtime statistics for the calculations with the CMS in default
setting are portrayed.
%
%
\begin{table}[htbp]
  \caption{\label{tab:runstats}Runtime statistics with
    CMS in default setting. The instance token comprises
    \emph{$\langle$number of rounds$\rangle$-$\langle$key
      token$\rangle$-$\langle$number of text pairs$\rangle$}.}
  \vspace{1ex}
  \begin{minipage}{\textwidth}
  \centering
  \begin{tabular}{c|r|r|r|r|r|r}
    \hline
    Instance\footnote{\textbf{k3}: 0123456789abcdef; \textbf{k4}: 0101010101010101; \textbf{k6}: b25286f7d3e7b3e1}
    & count & median & \multicolumn{2}{c|}{quartile} & mean & $\sigma$ [\%] \\
    \hline
    3-k4-12 & 25 & 2409.6 & 1243.8 & 3859.6 & 3821.9 & 126 \\
    3-k4-16 & 25 & 1569.8 & 993.6 & 3592.8 & 2700.8 & 98 \\
    3-k4-30 & 26 & 5281.6 & 4008.6 & 6221.2 & 5736.2 & 47 \\
    3-k3-12 & 17 & 167460.0 & 157160.3 & 175973.7 & 167778.0 & 21 \\
    3-k3-14 & 15 & 66963.4 & 47415.0 & 84625.9 & 74358.1 & 58 \\
    3-k3-18 & 11 & 26645.1 & 14093.4 & 40702.4 & 28979.0 & 71 \\
    3-k3-20 & 31 & 14073.1 & 6818.8 & 23159.6 & 27382.6 & 163 \\
    3-k3-20\footnote{Solution calculated with 21 threads.} & 8 & 56141.5 & 16057.7 & 105700.4 & 70640.1 & 98 \\
    3-k3-22 & 36 & 11788.1 & 7328.1 & 20468.9 & 17610.4 & 95 \\
    3-k3-30 & 4 & 34551.3 & 14286.1 & 54366.9 & 34101.8 & 70 \\
    3-k6s-20 & 50 & 23925.9 & 14190.5 & 34829.1 & 33958.6 & 112 \\
    3-k6s-22 & 33 & 26602.2 & 17846.1 & 45668.5 & 54138.4 & 157 \\
    3-k6s-24 & 51 & 9462.4 & 6100.0 & 17050.3 & 17113.6 & 111 \\
    3-k6s-30 & 22 & 64556.1 & 34625.1 & 96221.0 & 71600.6 & 76 \\
    \hline
  \end{tabular}
  \end{minipage}
\end{table}
%

%

\section{Empirical Parameter Optimization}
A dynamic code-analysis of the solver
preceded the practical parameter optimization phase, so as to
investigate how the solver-runtime consumption is distributed between
the various solver sub-processes and functions dependent on the external
parameter settings and the instance to solve. Again the dynamic
analysis results
are of statistical character, as the various solver modules and
functions are regularly called many times during the solver
runtime.
A previous static code analysis had provided associations
between external parameters and according parts of the code. The
question was, if dependent on the problem at hand, one could
possibly discover some optimal strategy of how to vary on parameters
influencing favorably the execution of the most time-consuming code parts, so as to effectively shorten the solution runtime.
For profiling the CMS program we used the GNU/Linux perf
tool\footnote{See for example \textsc{Wikipedia} perf (Linux)
  \url{https://en.wikipedia.org/wiki/Perf_(Linux)}.}. \emph{Flame
  Graphs} generated with the open source tool of the same name
developed by Brendan Gregg~\cite{FlameGraphs-2017} are utilized for
the visualization of the profiling results.
The code performance profiling has been carried out in both default
solver setting and with certain parameter settings other than default.
This delivered converging and unique results, as regards those parts of
the code causing the greatest CPU-load in case of all instances
implementing the attack on the 3 rounds encryption of the AES-64 model
cipher.
These results have been verified against different encryption
keys and different numbers of text pairs employed for the generation
of the problem instances and they appear to be stable.  82\% to 97\%
of the runtime the solver invests into its search routine (especially, the method \lstinline|propagate_any_order_fast()| of the propagation procedure) and this is
independent of whether an instance terminates during the observed
runtime or not.

Since the solution of the considered CNF instances typically takes a few thousand seconds, the profiling was performed only during a part of the program run. To insure a \emph{stable} result, we chose the measuring interval big enough, as well as the sampling frequency $f$ for perf high enough. One has to keep in mind that the performance profiling for CMS delivers a statistical statement. The reasons are: 1) the program runs of CMS are indeterministic; 2) in multi-thread modus of CMS, the profiling averages over all threads where the threads are computing independently of each other at different parts of the code; 3) the interval of the profiling measurement starts at different times of the running program.

\begin{figure}[htbp]
  \centering
  \includegraphics[scale = \inFGScale]{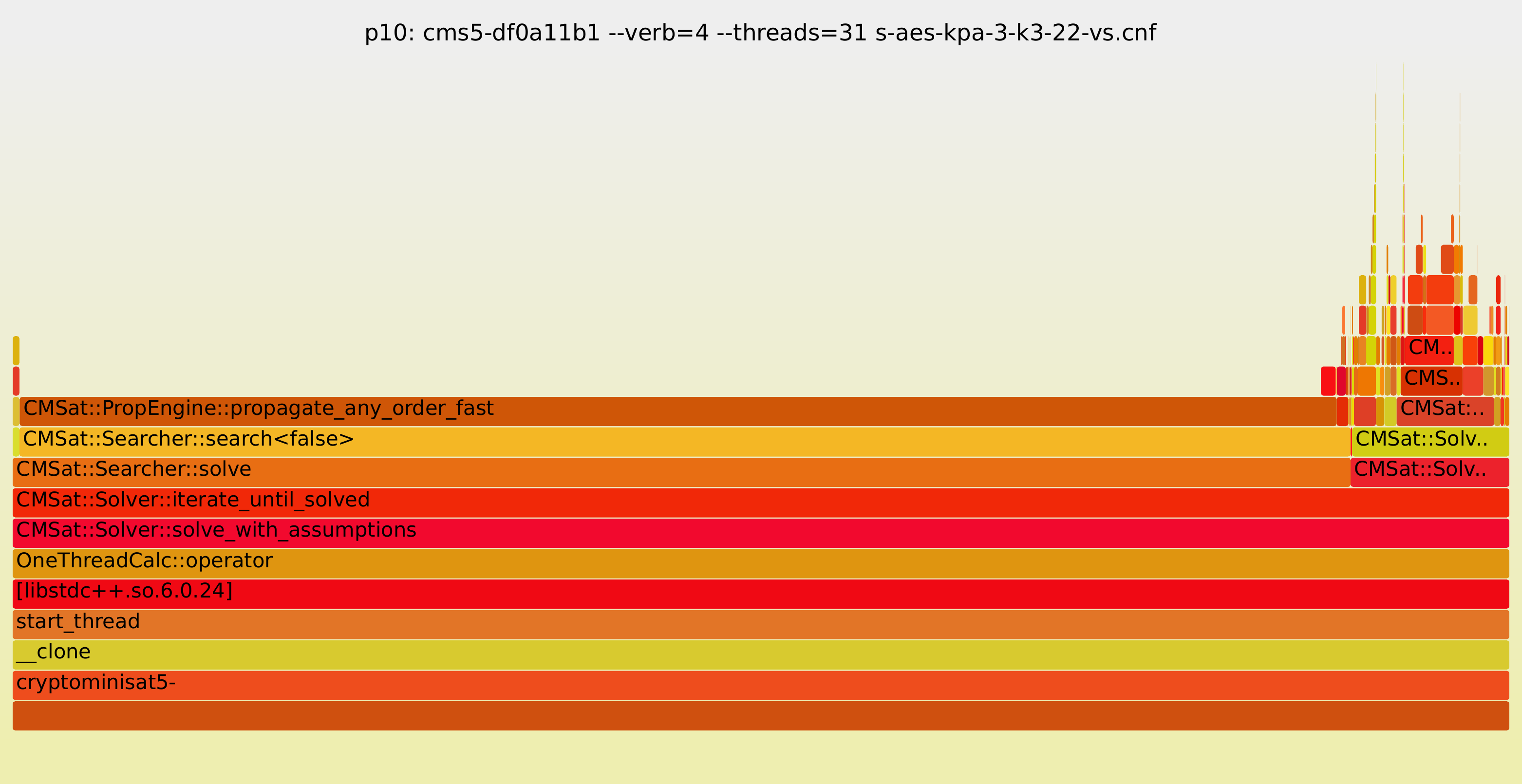}
  \caption[Flame Graph p10]
  {\label{f:FGp10}FlameGraph 3 rounds: 3-k3-22 with 31 threads, default parameter configuration, performance measuring with $f =$ 250 Hz for 300 s.}
\end{figure}

Exemplarily, a first profiling visualization is depicted in
Figure~\ref{f:FGp10}, where the relative runtime-shares which the most
used code paths of CMS consume, are depicted for the case of the medium hard to solve CNF
instance 3-k3-22. Apart from the verbosity level and the multi-thread
mode, CMS was here applied in default configuration.

Of interest is the variation of parameters influencing the whole of
the search process or such that influence the function
\lstinline|propagate_any_order_fast()| in a direct way.
In table~\ref{tab:SwitchCombis} there are listed parameters
influencing the solver's restart process, the choice of variables
and the setting of the so called \emph{glue values}.
\begin{table}[htbp]
  \caption{\label{tab:SwitchCombis}Explored parameter
    combinations and their abbreviations. Also set for all
    CMS runs: \lstinline|--verb=4 --threads=31 --comps=0|.
    sw3 und sw4 imply the usage of the here modified CMS.}
  \vspace{1ex}
  \begin{minipage}{\textwidth}
  \centering
  \begin{tabular}{l|l}
    \hline
    Abbreviation & Parameter Combinations \\
    \hline
    sw1 & \lstinline|--restart=geom --maple=1 --bva=0 --sync=30000| \\
    sw2 & \lstinline|--gluehist=30 --maple=1 --maxnummatrixes=8 --bva=0| \\
    sw6 & \lstinline|--restart=glue --gluecut0=4 --updateglueonprop=1| \\
    sw7 & \lstinline|--gluecut0=5 --gluecut1=7 --updateglueonprop=1| \\
    \hline
    sw3 & \lstinline|--restart=geom --maple=1 --cachesize=4096 --cachecutoff=3000| \\
    sw4 & \lstinline|--restart=glue --gluecut0=4 --updateglueonprop=1| \\
    \hline
  \end{tabular}
  \end{minipage}
\end{table}
The parameter \lstinline|--updateglueonprop| is the only one directly affecting
the function \lstinline|propagate_any_order_fast()|.
The last two parameter combinations, sw3 and sw4, indicate
calculations with a modified version of CMS. In the multi-thread mode,
CMS configures most threads slightly different than the command-line
settings in order to deliver good performance with the variety of CNF
instances to be solved in SAT solver competitions. To enhance the
influence of the here considered parameter combinations on the
solution of the examined cryptographic CNF instances, we changed the source
code so that the solver uses the same command-line settings for all threads.
Only the pseudo random number generator (PRNG) of each thread is seeded differently.
This change is effective, because it limits the solver's strategy to one exclusively examined configuration whereat the threads get diversified due to the usage of the PRNG and the asynchronous information exchange.

\begin{figure}[htbp]
  \centering
  \includegraphics[scale = \inFGScale]{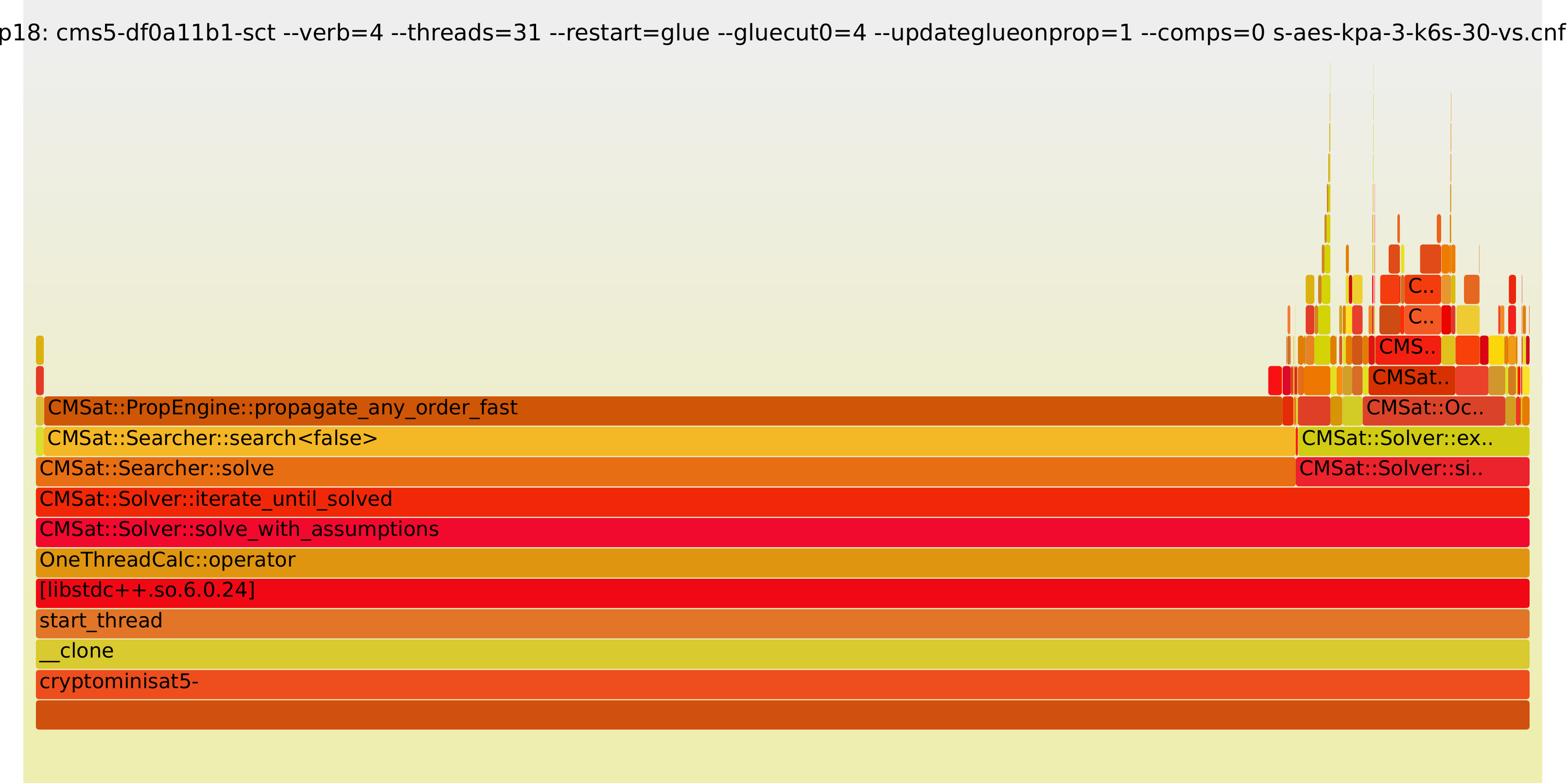}
  \caption[Flame Graph p18]
  {\label{f:FGp18}FlameGraph 3 rounds: 3-k6s-30 with 31 threads, parameter configuration sw4, performance measuring with $f =$ 250 Hz for 300 s.}
\end{figure}

In the FlameGraph-visualization of Figure~\ref{f:FGp18} there are
depicted the relative
runtime-shares which the most used code paths of CMS consume during
the calculations for the
\emph{harder} to solve CNF instance 3-k6s-30. In this case
the parameter combination sw4 was employed.
Here the solver spends a bit less time in the search as compared to
the the case shown in figure~\ref{f:FGp10}.

\begin{figure}[htbp]
  \centering
  \includegraphics[scale = \inBpScale]{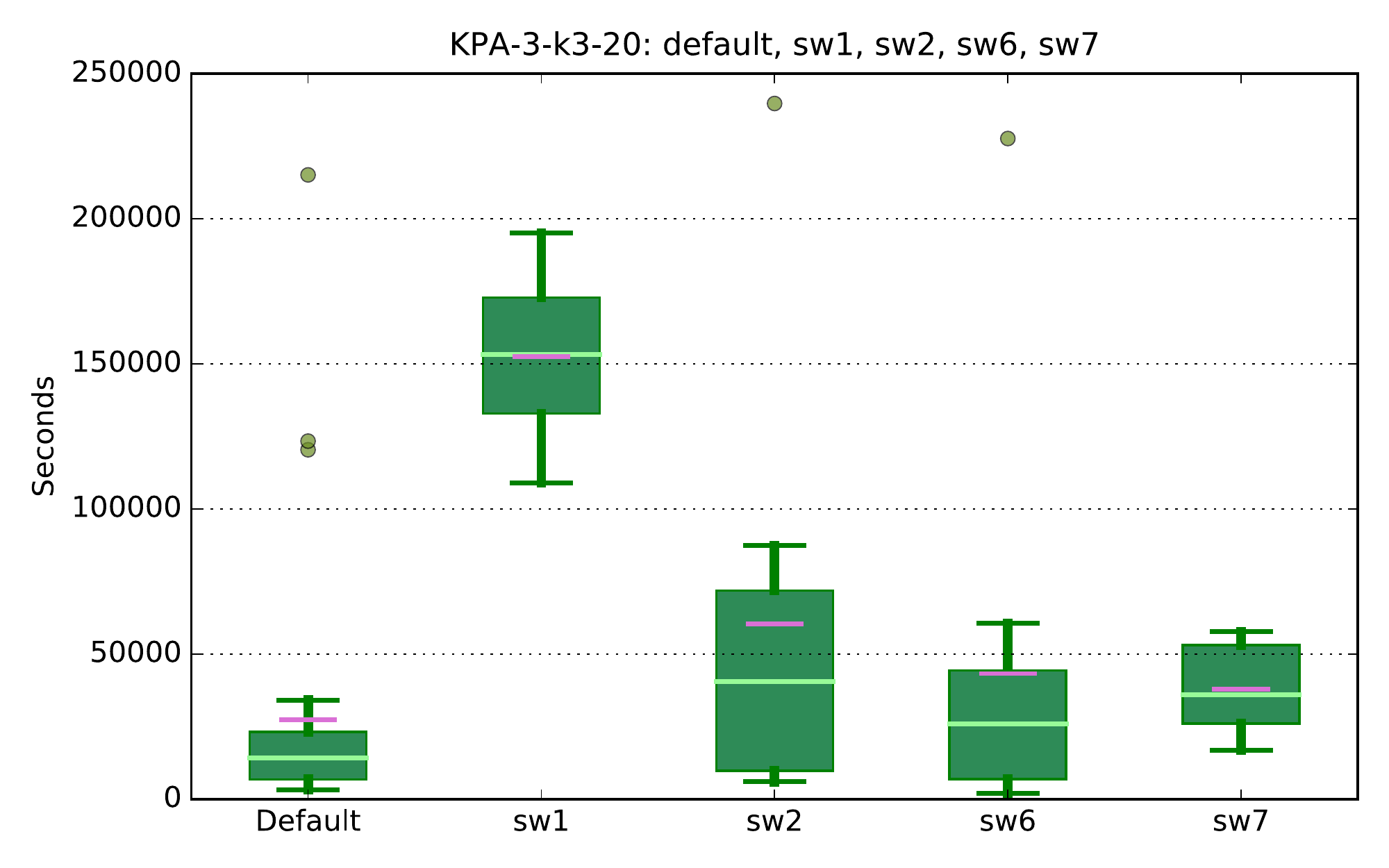}
  \caption[Boxplot 04]
  {\label{f:Boxplot04}3 rounds, key k3: 20 pair texts, parameter
    combinations: default, sw1, sw2, sw6, sw7.}
\end{figure}

\begin{figure}[htbp]
  \centering
  \includegraphics[scale = \inBpScale]{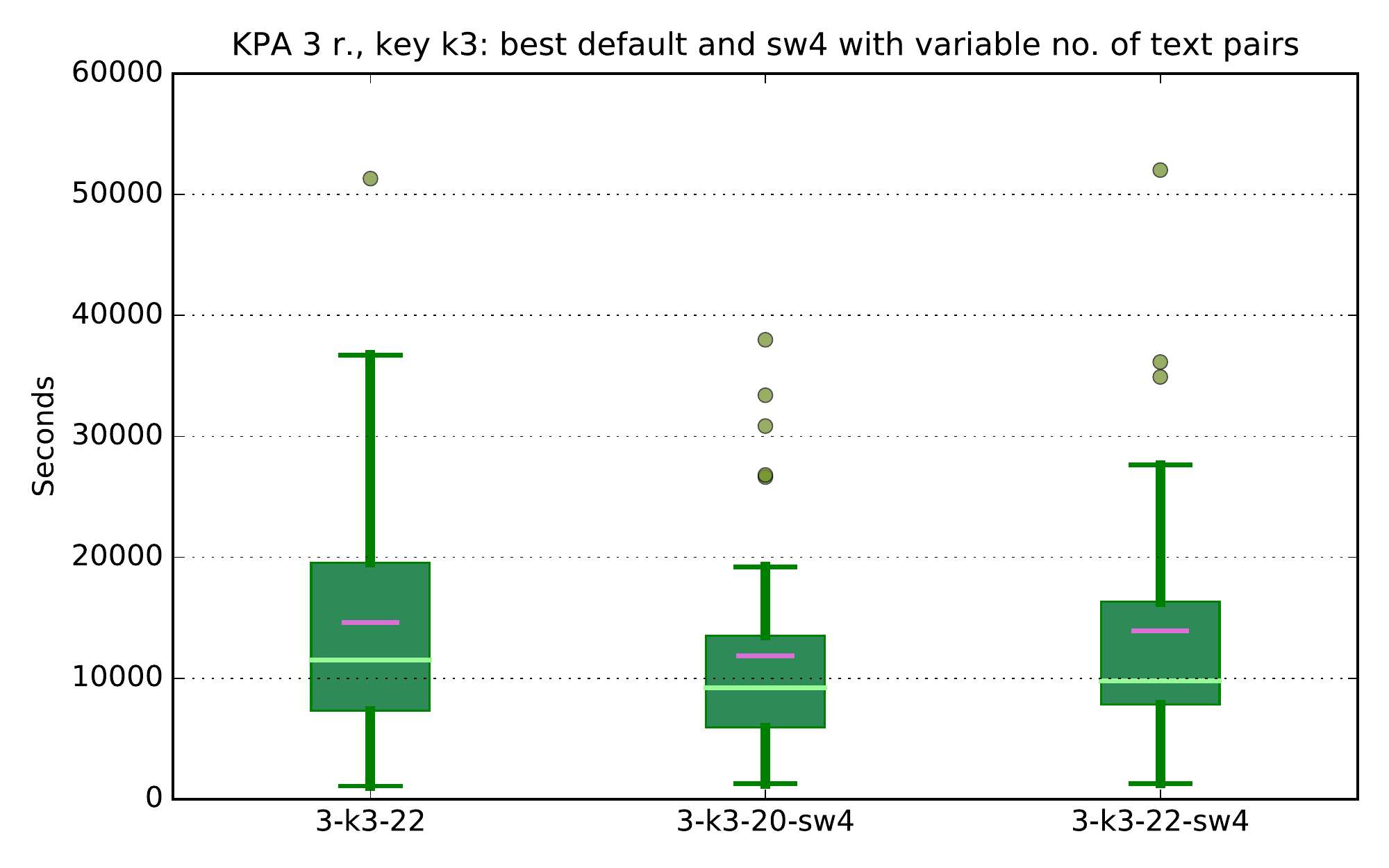}
  \caption[Boxplot 05a]
  {\label{f:Boxplot07a}3 rounds, key k3: best default, and sw4 with
    variable number of text pairs.}
\end{figure}

\begin{figure}[htbp]
  \centering
  \includegraphics[scale = \inBpScale]{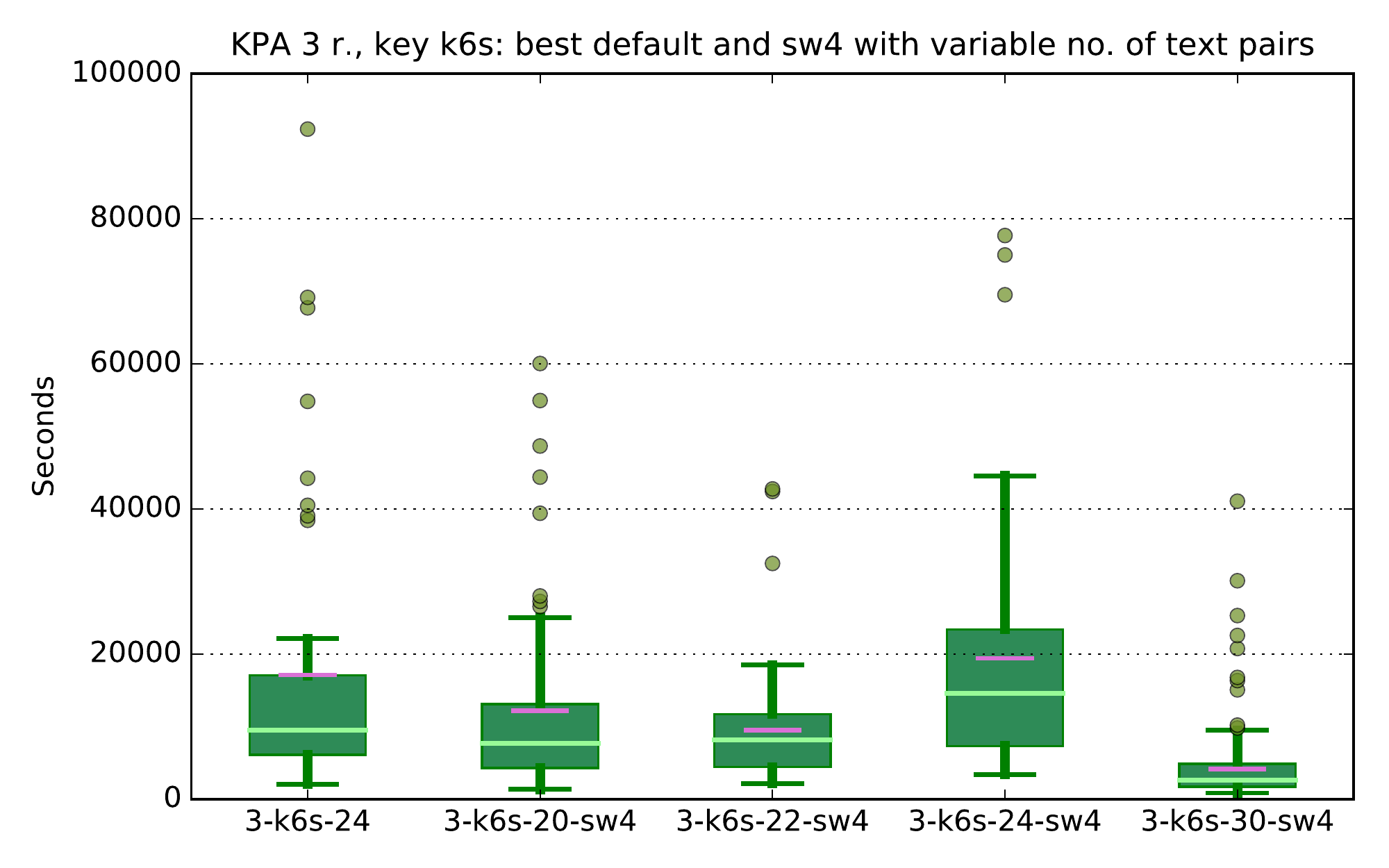}
  \caption[Boxplot 11a]
  {\label{f:Boxplot11a}3 rounds, key k6: best default, and sw4 with
    variable number of text pairs.}
\end{figure}

\begin{figure}[htbp]
  \centering
  \includegraphics[scale = \inBpScale]{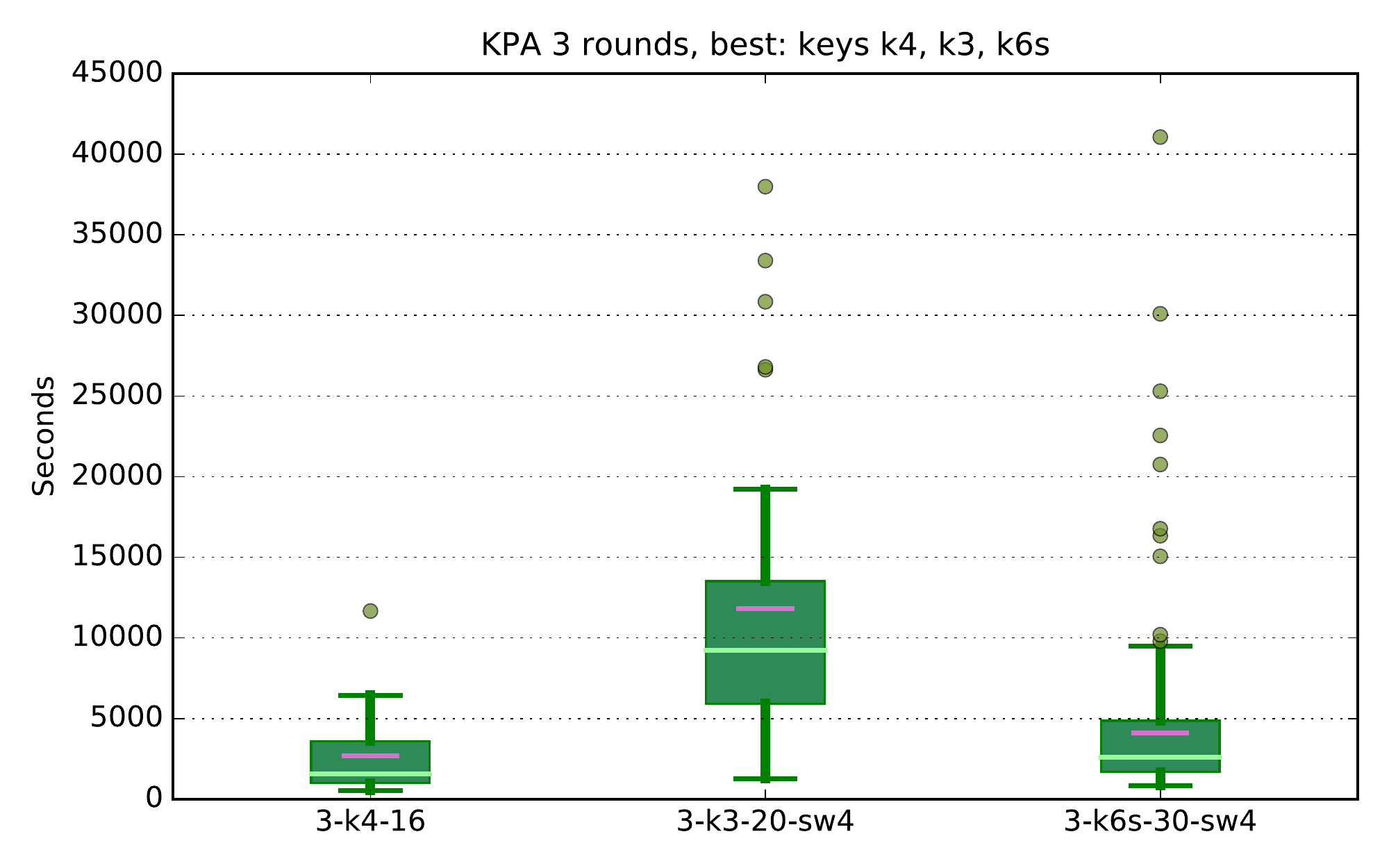}
  \caption[Boxplot 13a]
  {\label{f:Boxplot13a}3 rounds best for keys: k4, k3, k6s.}
\end{figure}

Because instances created with the \emph{insecure} key k4 could be solved
relatively fast with the CMS in default setting, only instances
created with the other two keys have been employed for the parameter
optimization tests. In Figure~\ref{f:Boxplot04} the solver runtime analysis for solving
instances created with the \emph{structured} key k3 is
demonstrated for different parameter combinations.
In the next Figure~\ref{f:Boxplot05a} the same instance as in Figure~\ref{f:Boxplot04} is tested with the parameter combinations: default, sw3 and sw4.
\begin{figure}[htbp]
  \centering
  \includegraphics[scale = \inBpScale]{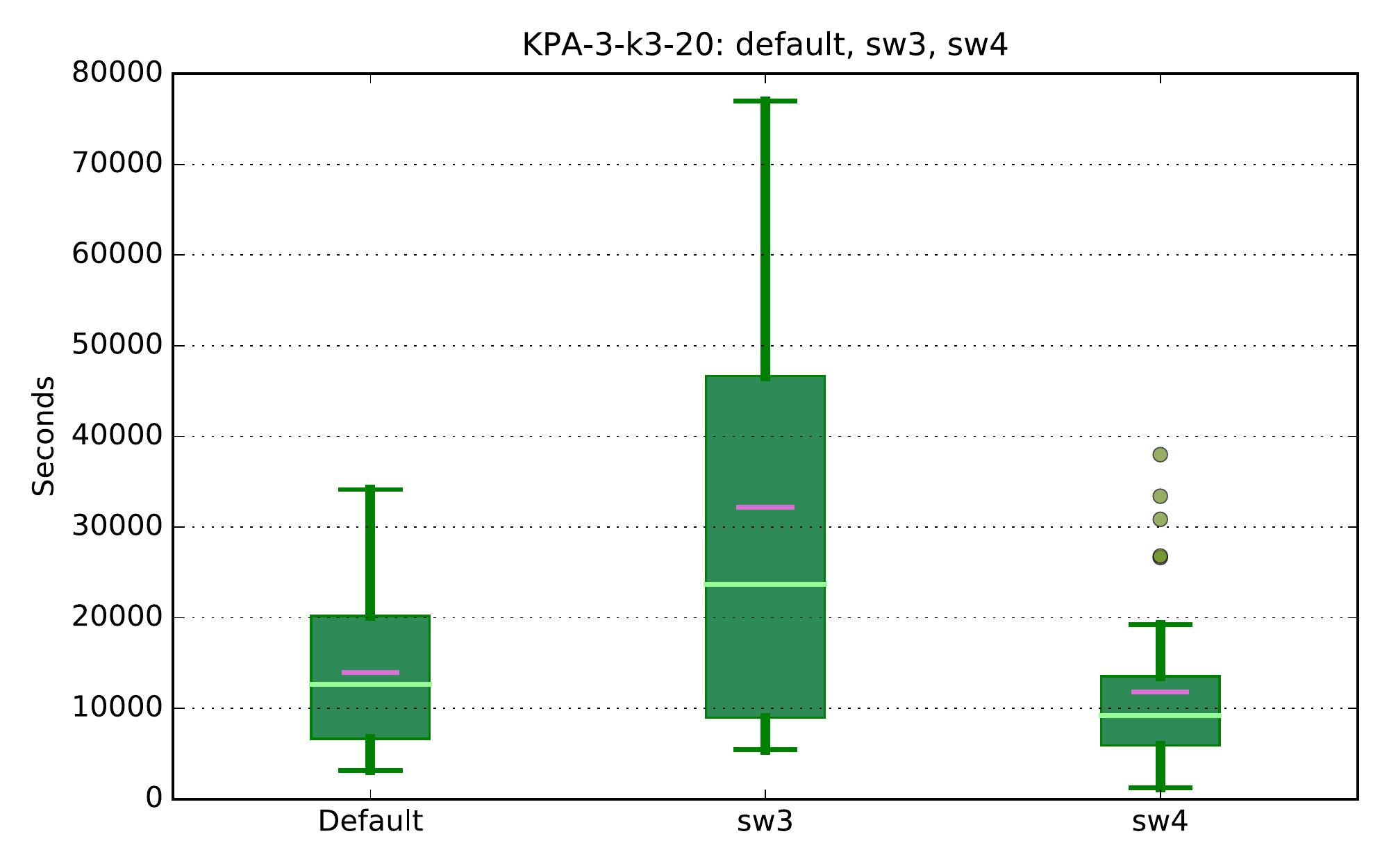}
  \caption[Boxplot 05a]
  {\label{f:Boxplot05a}3 rounds, key k3: 20 pair texts, parameter
    combinations: default, sw3 and sw4.}
\end{figure}
Evidently the choice of parameter combinations
has a considerable influence on the solution runtime of the solver.
An improvement in the solution time is registered with the
sw4 parameter combination as compared to the default setting. Changing
the number of text pairs we
test the stability of the sw4 runtime advantage, as
a function of the number of text pairs. We observe that this advantage
can get attenuated or
amplified, dependent on the chosen number of text pairs, see Figure~\ref{f:Boxplot07a}.
This suggests that the number of text pairs should also be observed as a problem optimization parameter.
There follows a series of runtime tests with sw4, solving instances created with the
\emph{secure} key k6. The results are exhibited in Figure~\ref{f:Boxplot11a}.
In Figure~\ref{f:Boxplot11a} we see that the instance created with 30
text pairs allows the faster reconstruction of the key k6 when the
parameter combination sw4 is applied. Notably, even the upper quartile
of the optimized best solver runtime, for the instance with the 30
text pairs, lies underneath the lower quartiles of the runtimes of all
other instances, thus establishing the unambiguity of this result. The
following Boxplot in Figure~\ref{f:Boxplot13a} depicts the comparison
between the best results for all three different keys.
It seems that for every key and independent of its security quality,
there exists a combination of a number of text pairs and a CMS configuration to
find the solution with a statistical median lying well below the 10000 seconds
runtime limit for the solver. This is a significant result because it indicates that
a secure key might not necessarily offer better protection
against solving the here discussed instances.
In table~\ref{tab:runstats2} the solver runtime statistics for the practically
optimized parameter settings are presented.
%
%
\begin{table}[htbp]
  \caption{\label{tab:runstats2}Runtime statistics
    with various parameter combinations for CMS. Instance tokens
    comprise \emph{$\langle$no.\ of rounds$\rangle$-$\langle$key
      token$\rangle$-$\langle$no.\ of text pairs$\rangle$}.}
  \vspace{1ex}
  \begin{minipage}{\textwidth}
  \centering
  \begin{tabular}{c|r|r|r|r|r|r}
    \hline
    Instance\footnote{\textbf{k3}: 0123456789abcdef; \textbf{k6}: b25286f7d3e7b3e1}
    & count & median & \multicolumn{2}{c|}{quartile} & mean & $\sigma$ [\%] \\
    \hline
    3-k3-20-sw1 & 4 & 153149.7 & 132870.1 & 172872.3 & 152592.8 & 24 \\
    3-k3-20-sw2 & 9 & 40596.2 & 9753.6 & 71882.5 & 60456.5 & 121 \\
    3-k3-20-sw3 & 12 & 23689.6 & 8963.1 & 46646.4 & 32149.3 & 87 \\
     3-k3-20-sw4 & 35 & 9227.2 & 5919.9 & 13522.9 & 11828.3 & 79 \\
    3-k3-20-sw6 & 11 & 25895.9 & 6922.7 & 44363.8 & 43264.8 & 148 \\
    3-k3-20-sw7 & 8 & 36036.9 & 26012.9 & 53067.9 & 37975.0 & 42 \\
    3-k3-22-sw4 & 35 & 9761.5 & 7819.0 & 16320.9 & 13895.4 & 79 \\
    3-k6s-20-sw4 & 60 & 7575.2 & 4225.7 & 12310.2 & 10686.0 & 95 \\
    3-k6s-22-sw4 & 74 & 8142.4 & 4410.4 & 11720.8 & 9535.7 & 80 \\
    3-k6s-24-sw4 & 59 & 14547.9 & 7312.2 & 23404.8 & 19408.2 & 88 \\
    3-k6s-30-sw4 & 197 & 2581.0 & 1687.8 & 4857.5 & 4110.9 & 116 \\
    \hline
  \end{tabular}
  \end{minipage}
\end{table}
%

%

\section{Automatic Algorithm Configuration (AAC)}

The adaption of SAT solver configurations to a specific type of
instances or instance classes is a common practice employed by many
developers of such programs participating in the international SAT
solver competition.\footnote{International SAT Solver Competitions
  \url{http://www.satcompetition.org}.} Meanwhile, also computer tools
for the automatic algorithm configuration are available. Such programs
even participate in the \emph{Configurable SAT Solver Challenge (CSSC)}, organized by F.~Hutter et al.~\cite{Hutter2017:CSSC}. Led by the results of the CSSC 2016 we chose the tool SMACv3, developed by M.~Lindauer, F.~Hutter et al.~\cite{SMACv3-2017,Hutter2011:SMO-SMAC} at the Universities of Freiburg and British Columbia, to further optimize the parameter settings of CMS for cryptographic CNF instances.

In order to apply SMACv3 with CMS, we set up the required Python
environment and implemented a \emph{Target Algorithm Evaluator} (TAE),
a Python wrapper around CMS version 5.0.1, for the SMACv3 optimization
API. We defined the legal ranges of the parameters to be optimized by
SMACv3 by setting up a \emph{Parameter Configuration Space}
(PCS). From the PCS, SMACv3 chooses parameter combinations and calls
the TAE with it. The resulting runtime of a CMS computation returned
via the TAE is evaluated by SMACv3 for the parameter optimization and
for further calls of the TAE.
SMACv3 is able to deal with indeterministic runtimes by repeatedly calling CMS with the same parameter configuration and evaluating an estimator for the runtime.
We performed the optimization by using SMACv3 in parallel mode.

\begin{figure}[htbp]
  \centering
  \includegraphics[scale = \inBpScale]{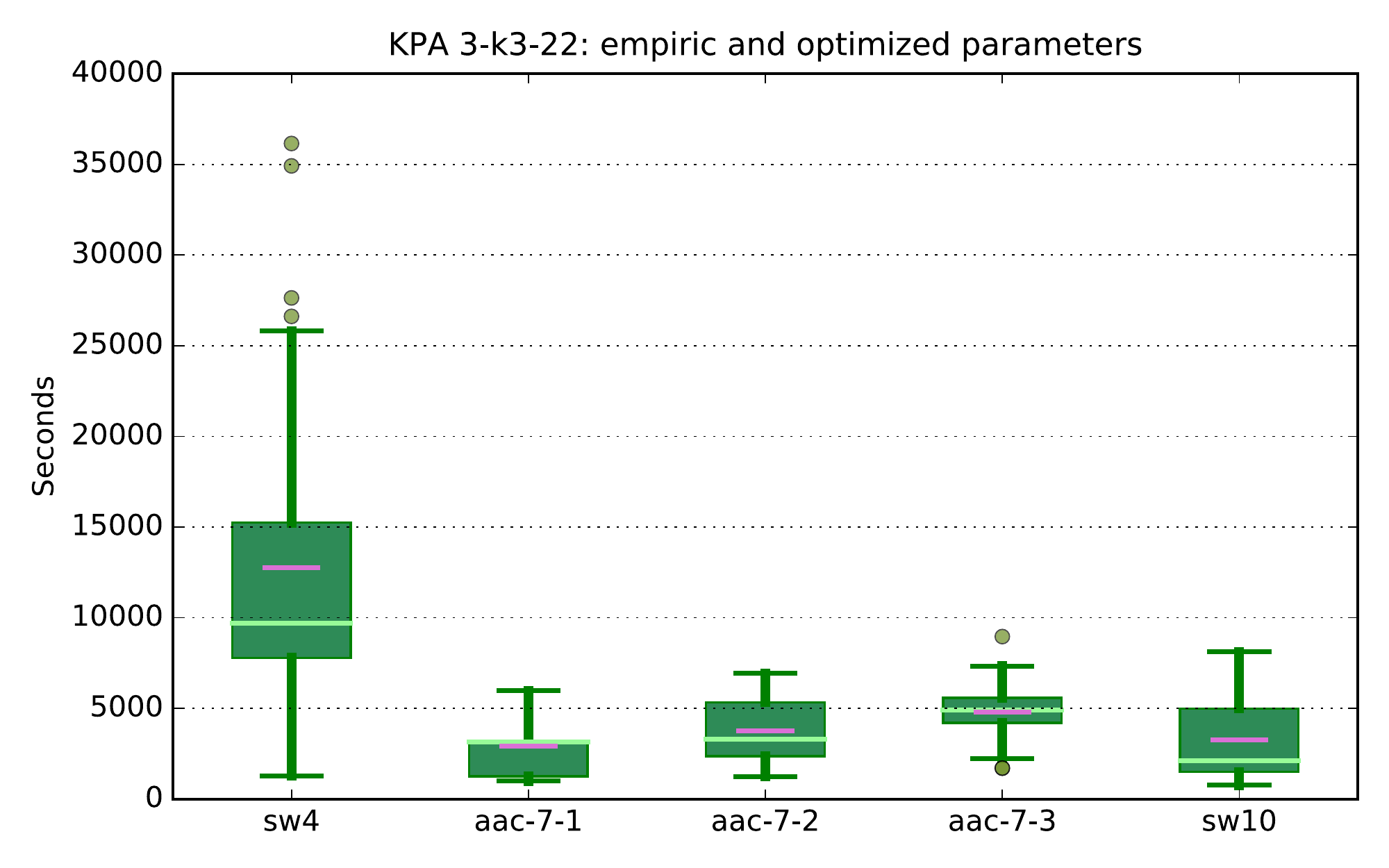}
  \caption[Boxplot aac-7]
  {\label{f:BoxplotAAC7}3 rounds, key k3, 22 text pairs: sw4, 3 optimizer \emph{incumbents}, and sw10.}
\end{figure}

\begin{figure}[htbp]
  \centering
  \includegraphics[scale = \inBpScale]{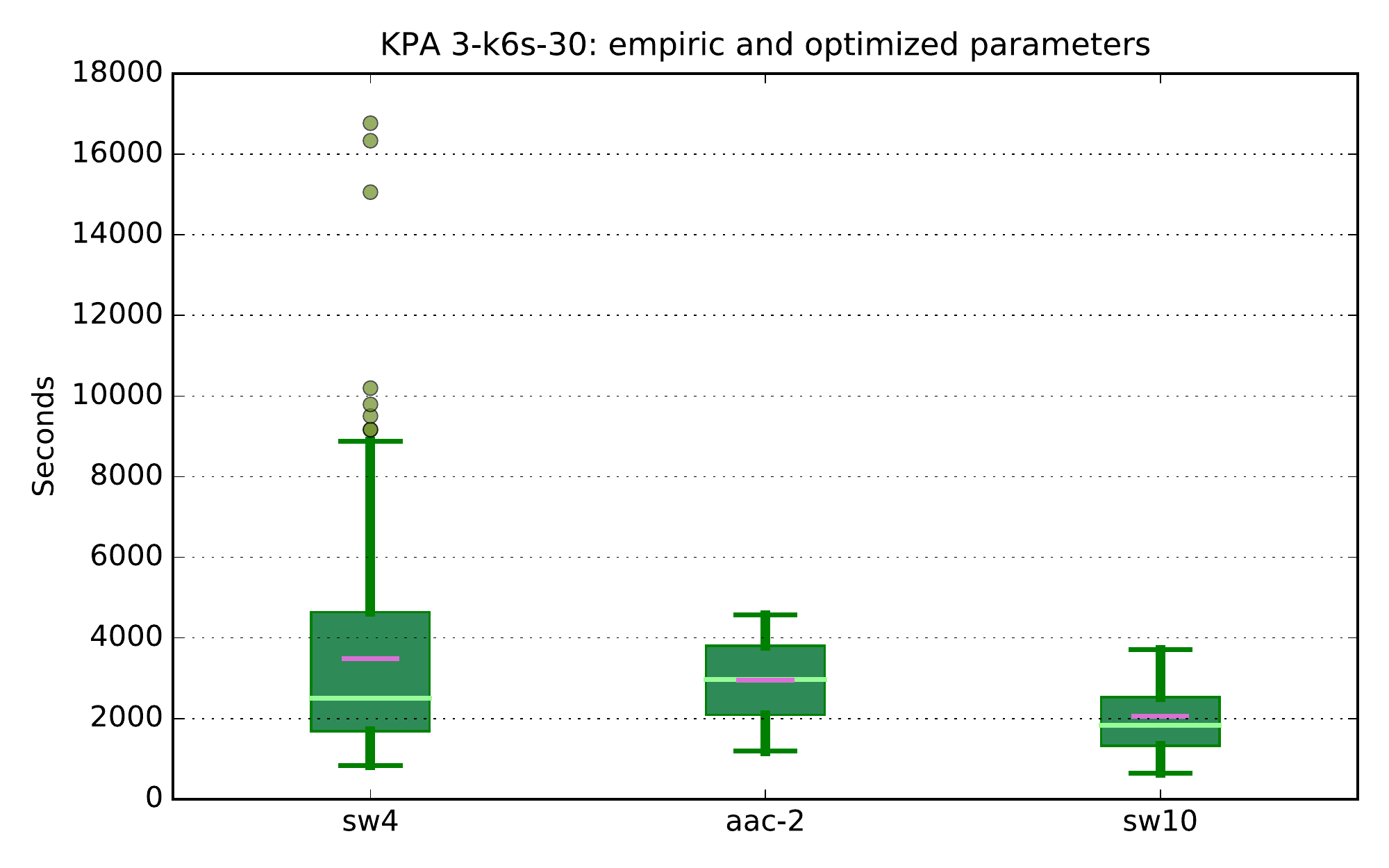}
  \caption[Boxplot aac-2]
  {\label{f:BoxplotAAC2}3 rounds, key k6s, 30 text pairs: sw4, an optimizer \emph{incumbent}, and sw10.}
\end{figure}

All parameter optimization runs were performed with the modified
version of the CMS code with alike configuration for all
threads. Starting from our empirical configuration results, we
confined the parameter optimization to few CMS parameters, a typical
PCS file of ours looks like follows:
\begin{lstlisting}
  # Restart options
  gluehist [40, 250] [50]i
  # Red clause removal
  gluecut0 [1, 6] [3]i
  gluecut1 [5, 9] [5]i
  adjustglue [0.3, 0.9] [0.7]
  # Variable branching options
  freq {0.0, 0.1, 0.2, 0.3, 0.4} [0.0]
\end{lstlisting}

We performed optimizations for the CNF instances 3k3-22 and 3-k6s-30,
independently. In Figures~\ref{f:BoxplotAAC7} and \ref{f:BoxplotAAC2}
we compare some intermediate states of the optimization (called
\emph{incumbents} by SMACv3) with the empirical best configuration
sw4.
From the results of several optimization runs (each computing
nearly a week) we excerpted a configuration which reduces the median
runtime for both instances. Compared to the empirical parameter setting
sw4, the new best parameter combination sw10 sets additionally
\lstinline|--gluecut1=7 --gluehist 45|. The Boxplots of the runtime analysis
for this parameter setting are also shown in Figures~\ref{f:BoxplotAAC7} and \ref{f:BoxplotAAC2}.
The numerical values of the Boxplot estimators are recapitulated in
table~\ref{tab:runstats3}.
%
%
\begin{table}[htbp]
  \caption{\label{tab:runstats3}Runtime statistics
    of better parameter combinations for CMS. Instance tokens comprise
    \emph{$\langle$no.\ of rounds$\rangle$-$\langle$key
      token$\rangle$-$\langle$no.\ of text pairs$\rangle$}.
    Times are in seconds for the solving-thread.}
  \vspace{1ex}
  \begin{minipage}{\textwidth}
  \centering
  \begin{tabular}{c|r|r|r|r|r|r}
    \hline
    Instance\footnote{\textbf{k3}: 0123456789abcdef; \textbf{k6}: b25286f7d3e7b3e1}
    & count & median & \multicolumn{2}{c|}{quartile} & mean & $\sigma$ [\%] \\
    \hline
    3-k3-22-sw4 & 35 & 9761.5 & 7819.0 & 16320.9 & 13895.4 & 79 \\
    3-k3-22-aac-7-1 & 5 & 3134.1 & 1256.4 & 3212.2 & 2913.7 & 68 \\
    3-k3-22-aac-7-2 & 10 & 3291.5 & 2360.9 & 5333.9 & 3767.1 & 55 \\
    3-k3-22-aac-7-3 & 25 & 4925.2 & 4201.4 & 5574.6 & 4792.0 & 35 \\
    3-k3-22-sw10 & 15 & 2127.3 & 1503.9 & 4993.0 & 3262.4 & 72 \\
    \hline
    3-k6s-30-sw4 & 197 & 2581.0 & 1687.8 & 4857.5 & 4110.9 & 116 \\
    3-k6s-30-aac-2 & 8 & 2966.6 & 2085.9 & 3807.3 & 2956.1 & 41 \\
    3-k6s-30-sw10 & 15 & 1830.1 & 1321.5 & 2526.4 & 2057.4 & 46 \\
    \hline
  \end{tabular}
  \end{minipage}
\end{table}
%

%
The runtime of the CNF instance 3-k6s-30 could be improved by almost 30\% and that of the instance 3-k3-22 even by nearly 80\% in median.

\section{Conclusions and Work in Progress}
In this paper we describe the steps taken in studying the influence of
various CMS configurations on the SAT solver's performance in trying
to find solutions for cryptographic instances representing algebraic
known-plaintext attacks on the 3 rounds small AES-64 model cipher.
The static and dynamic analysis of CMS has pointed to the most computationally
intensive parts, which in turn motivated variations of certain
configuration parameters expected to influence
the execution time of mainly these parts of the code. We also modified the
source code in a way that enhances configuration changes and produces
clearer results.
We performed statistical runtime analysis of a plethora of results
created in both default and other solver configurations which enabled
us to identify solver configurations that solve the here discussed
instances and thus fully recover the 64-bit key in time intervals
underneath an hour (real time).\footnote{The solution is found by one thread in this time, but not without exchange with the other threads also searching a solution in this time. In a sense one could say: CPU solution time of the winning thread.} This result is independent of the
security quality of the key.
By means of an Automatic Algorithm Configuration (AAC) we could
even improve the previous best runtimes achieved with
empirically decided configurations. Since CMS has many more
parameters than varied here, it potentially
offers possibilities for further configuration optimization. Therefore we
intend to expand our efforts in this direction with AAC.
The universality of validity of the here elaborated configurations
has still to get verified. This demands the creation of many more
instances and running of many more tests. Also experiments with
other types (e.g. \emph{dense}) of CNF-instances have to be considered.
Perhaps, when
appropriately tuned, CMS can solve even bigger cryptographic problems.
Increasing the number of
rounds means having to handle CNF-instances with many more variables and many more
millions of constraints in comparison to the here handled
problems. A worst case complexity would be soon formally reached.
However, \emph{in the SAT
  solver context, worst case complexity has no explanatory or
  predictive power}~\cite{Ganesh2015:SATcomplexity}.

%
\bibliographystyle{plain}
\bibliography{pos18}
\label{sec:references}
\end{document}